# Microwave impedance microscopy imaging of acoustic topological edge mode on a patterned substrate

Y. Nii,[1,2*] and Y. Onose[1]

1. Institute for Materials Research, Tohoku University, Sendai 980-8577, Japan

2. PRESTO, Japan Science and Technology Agency (JST), Kawaguchi 332-0012, Japan

We have studied acoustic topological edge modes in a honeycomb phononic crystal composed of metallic nano-pillars on a $LiNbO_3$ substrate. Acoustic band calculations show that the topological surface acoustic wave (SAW) mode inhabits the edge of the honeycomb phononic crystal in spite of the hybridization with the internal acoustic modes of the substrate. Pulse-type microwave impedance microscopy realized clear visualization of the gigahertz topological edge mode between two mutually inverted topological phononic crystals. A frequency-dependent image showed that the edge mode evolves as the bulk SAW modes are suppressed owing to the energy gap formation, consistent with the topological nature. The realization of a topological waveguide in a simple pillar structure on a substrate might pave a new path to the development of topological SAW devices.

**MAIN TEXT**

Since the discovery of the quantum Hall effect (QHE) [1], the topological aspect of electronic states has attracted much attention [2-4]. One of the most prominent properties of topological electronic states is the formation of edge or surface states. In a two (three)-dimensional topological crystal, for example, the edge (surface) is conducting even when the Fermi energy is located in an energy gap. The anomalous properties, including robust one-way transport along the edge (surface), are thought to be the hallmark of topological electronic states. The concept of topology is not restricted to the electronic states but is common to other Bloch states, such as photons [5], magnons [6], and phonons [7,8]. For these bosonic quantum waves, the topological edge state should work as a low dissipative waveguide, offering significant potential for applications.

Of particular importance may be the phonon since it carries sound and heat [9]. The control of heat is one of the most important issues for resolving energy-related problems. Sound waves or higher frequency acoustic waves are also widely utilized in modern society. Among various acoustic devices, surface acoustic wave (SAW) devices have been important components of telecommunication equipment [10] and chemical/bio-sensors [11], and the controllability of acoustic waves contributes to their further development. For these purposes, in the field of phononics, intensive studies have been performed for realizing various properties, such as negative refraction [12], cloaking [13], and rectification [14, 15]. In particular, artificial periodic structures with a period comparable with the wavelength, termed a phononic crystal (PnC), have been developed to realize such properties. This approach can be applied to research on topological phononic properties.

Among the wide phononic spectrum ranging from kHz sound waves up to THz thermal phonons, the most prominent progress in topological research can be seen in airborne sound waves in the kHz range [16-21]. Since the wavelength of kHz sound waves is relatively long, it is possible to fabricate topological phononic crystals with centimeter-scale periodicity. Consequently, various topological acoustic phases have been quickly realized in the kHz region, including the Zak phase [16], the QHE [17], the quantum spin Hall effect [18], the valley Hall effect [19], the higher-order topological state [20], and Weyl semimetal states [21]. Compared with these low-frequency airborne sound waves, only a few experiments have succeeded in realizing topological acoustic states operating at much higher frequencies, such as MHz [22-25] and GHz [26], partly due to the technical difficulty of fabricating smaller topological devices and observing topological transport. However, it seems important to realize topological acoustic properties at these high frequencies, especially in the GHz range, since the frequencies are compatible with commercial telecommunication frequencies. Quite recently, with the use of microwave impedance microscopy (MIM), Zhang *et al.* observed a topological edge state of a 1 GHz acoustic wave in a two-dimensional PnC fabricated on a free-standing membrane [26]. However, the fabrication process of such a structure is cumbersome for realizing practical devices. Here, we reveal that a patterned piezoelectric substrate hosts a similar topological waveguide of a SAW by means of acoustic band calculations and MIM. The simplified fabrication method of such a topological waveguide would be more applicable to practical devices.

To realize the topological edge state of acoustic waves, we fabricated periodically arranged nano-pillars on a Z-cut LiNbO$_3$ substrate forming honeycomb lattices with a lattice constant $L$ of 840 nm, as shown in Figs. 1(a) and 1(b). Their unit cell consisted of two different Au cylinder-like pillars with the same height $h$ = 158 nm and substrate. The average diameter and asymmetric factor, $\alpha = (d_m - d_n)/(d_m + d_n)$, of the bottom diameters, $d_m$ and $d_n$, for these pillars were 273 nm and 0.37, respectively. The diameters at the upper surfaces of both pillars were 85% of the bottom ones. In electronic systems, a honeycomb lattice with two inequivalent atoms in a unit cell is known to show the quantum valley Hall effect (QVHE) [27]. In the QVHE, nontrivial Berry curvature diverges around the pair of energy extrema in momentum space, denoted as K and K' valleys. Although the total Berry curvature must be zero owing to the preserved time-reversal symmetry, its integration around each valley becomes nonzero, producing valley-dependent Chern numbers (called valley Chern numbers). Then, topological valley transport may emerge at the interface of materials having valley Chern numbers of opposite signs. The question here is whether the SAW version of the QVHE is realized in the honeycomb latticed pillars on a substrate. This is not so simple to answer because the internal acoustic wave mode of the substrate may be hybridized with the topological surface mode. To answer this question, we performed acoustic band calculations for the system described above [Fig. 1(b)] using the finite element method. To simplify the calculation, we assumed that the thickness of the substrate was four times as large as the lattice constant of the honeycomb lattice $L$, while the thickness was much larger (0.5 mm) in the experiment. Details of the calculation are shown in the Supplemental Materials. Figure 1(c) shows the obtained acoustic band for $\alpha = 0$ and $\alpha = 0.37$. In order to distinguish a SAW localized on the surface and a bulk acoustic wave (BAW) extending over the interior of the substrate, we evaluated for the case of $\alpha = 0.37$ how much strain energy is concentrated on the surface. The dark-red (white-yellow) color indicates that elastic energy was more localized at the surface (extended over the bulk), meaning that the eigenmode had a SAW (BAW) character. Because of the finite thickness of the substrate, many standing wave modes with bulk-like nature were discerned close to the Γ point at finite frequencies. As the thickness increased, the number of standing wave modes increased but all the bulk-like modes settled in the colored region irrespective of the substrate thickness. On the other hand, the surface-like flat modes were discerned at around 1.4 GHz and 2.0 GHz, originating from localized pillar resonances, as previously confirmed in various pillar-type PnCs [24, 25, 28]. Most important here are the dispersive surface modes around the K point. For $\alpha = 0$, the dispersive SAW modes showed band crossing so that a Dirac point appearred at the K point around 2.4 GHz. This is consistent with previous reports in the MHz range [28]. When the asymmetry $\alpha$ was introduced, the two-fold degeneracy at the K point was lifted, and the SAW band gap emerged, indicating the topological valley Hall insulating state of the SAW. Importantly, these SAW modes around the K valley were well separated from the BAW modes in the colored area. Therefore, the topological nature of the two-dimensional honeycomb PnC was preserved even in the presence of the substrate. To scrutinize the topological transition, the $\alpha$ dependences of acoustic mode frequencies at the K point are displayed in Fig. 1(d). In the course of the reversal of $\alpha$, the bandgap at the K valley first closed and then reopened, corresponding to the reversal of valley Chern numbers. The upper and lower branches had clockwise or counter-clockwise rotational strain motion. As depicted by red or blue colors in the inset of Fig. 1(d), these two rotational states are analogous to the electronic states with pseudospins of opposite sign.

To directly visualize a topological SAW mode at gigahertz frequency with sub-micron spatial resolution, we developed a pulse-type MIM imaging system, as shown in Fig. 2(a). It is based on a commercial atomic force microscopy (AFM) system (EDU-AFM, Thorlabs) and a conductive cantilever (12Pt-400B, Rocky Mountain Nanotechnology). Microwave electronics consist of a microwave generator (MG3692C, Anritsu), a pulse generator (DG645, Stanford Research Systems), an IQ-mixer (MMIQ-0218LXPC, Marki Microwave), and a boxcar integrator (SR250, Stanford Research Systems). The unique feature different from conventional MIM systems [26, 29-31] is to use pulse modulated microwaves with a pulse duration of 150 ns. In this system, one can separate SAW and electromagnetic crosstalk in the time domain. Since the distance from the interdigital transducer (IDT) to the PnC was about 1.3 mm, SAW signals arrived at the PnC 0.3 μs late from electromagnetic crosstalk. The time-delayed SAW signal was selectively detected by means of the boxcar integrator, and was digitized by a data acquisition (DAQ) system (NI USB-6361). An IDT and topological PnC were fabricated on a Z-cut $LiNbO_3$ surface by using a standard electron beam lithography and lift-off technique. In order to excite a SAW over a wide frequency range between 2.2 GHz and 2.55 GHz, a chirped IDT was fabricated, as schematically shown in Fig. 2(a), in which the period of the IDT was continuously varied from 1.48 μm to 1.62 μm. To realize a topological SAW waveguide, we prepared two honeycomb lattices with valley Chern numbers of opposite signs, shown in Figs. 2(b) and 2(c), which are hereafter denoted as A and B lattices, respectively. The boundary between the upper A lattice and the lower B lattice was located at the center of the SAW propagation area, as schematically shown in the inset of Fig. 2(a). Figures 2 (d) and 2(e) show typical AFM and MIM images around the AB interface. These AFM and MIM images were obtained simultaneously under SAW excitation at 2.22 GHz. On the lefthand side where there was no PnC, the periodic intensity modulation existing only in the MIM image corresponds to the SAW launched from the left IDT. This is also confirmed by line profiles of AFM and MIM displayed in Figs. 2(f) and 2(g), respectively. The average height of the pillars measured by AFM was 153 nm, which is close to the designed total thickness of 158 nm.

To demonstrate the topological SAW edge mode, we show MIM images in the vicinity of the AB interface [see Fig. 3(a)] at various frequencies between 2.22 GHz and 2.52 GHz in Figs. 3(b)-3(h). At 2.22 GHz and 2.30 GHz, the SAW propagating from the left side readily penetrated the PnC, showing the strong contrast caused by the PnC structure. The large intensity variation in the PnC seems to have been caused by interference with the reflected SAW at the PnC boundary and the scattered SAW due to imperfections. On the other hand, at 2.52 GHz, the SAW intensity was quite suppressed in the PnC, whereas the wavefront of the SAW was clearly seen before the PnC boundary. Between 2.35 GHz and 2.45 GHz, SAW propagation was discerned only around the boundary of the A-B lattices, which is the topological edge mode as discussed below.

To clarify the nature of the boundary mode, we examine the vertical position-dependent SAW intensity in the PnC shown in Fig. 4(a), compared with the band calculation around the K valley shown in Fig. 4(b). We obtained the intensity by Fourier transformation (FT) of the MIM image. The details of the FT analysis are shown in the Supplemental Material. In the band calculation below 2.3 GHz, there exists a *bulk* surface acoustic mode around the K point. For this reason, we observed a large SAW intensity in the interior of A and B lattices in this frequency range. The spatial modulation of the intensity is caused by the aforementioned scattering and interference effect. Above 2.35 GHz, the *bulk* surface acoustic state is gapped but the edge state should emerge according to the band calculation. This is in agreement with the experimental observation: The SAW intensity is discernible only around the interface. The edge mode intensity shows a maximum around 2.39 GHz, which seems to correspond to the Drac point for $\alpha = 0$. In the higher frequency range, the upper *bulk* surface acoustic band shows up, but the dispersion relation is almost flat. Therefore, the SAW intensity is localized at the boundary of the PnC and suppressed inside the PnC regardless of the location. Faint contrast reflecting the PnC may originate from some residual electromagnetic crosstalk. Figure 4(c) summarizes the edge and bulk mode intensities as a function of frequency. The intensity of the edge mode was deduced from Gaussian fitting to the position-dependent SAW intensity around the interface. The bulk mode intensity was evaluated from the FT intensity inside the PnC normalized by that outside the PnC. While the bulk mode intensity is suppressed above 2.35 GHz, the edge mode intensity rapidly increases around 2.35 GHz and shows a peak around 2.39 GHz. A small difference in the bandgap frequency between calculation and experiment may be attributable to the derivation of some material parameters used in the calculation and the fabricated PnC (see Supplemental Material). The contrastive frequency dependence of the bulk and edge intensities ensures the validity of the topological edge state.

In conclusion, we have demonstrated a topological SAW mode around 2.4 GHz in a honeycomb lattice made of metallic pillars on a $LiNbO_3$ piezoelectric substrate. With the use of a pulsed MIM system, we visualized the topological edge mode. The frequency dependence of the edge mode intensity is contrastive with that of the *bulk* SAW mode, which ensures the topological nature. While a similar topological edge state was previously realized in a purely two-dimensional system, the present result shows that the topological edge mode traveling along the surface is robust against hybridization with the interior BAW modes of the substrate. The method of topological patterning on a substrate seems quite useful for implementing topological waveguides and other functionalities into practical acoustic devices working at GHz frequencies.

**Acknowledgments**

We are grateful for K. Lai, D. H. Lee, and L. Zheng for the technical instruction of MIM imaging, and also grateful for T. Seki for technical advice for fabricating topological PnC. This work is supported by JSPS KAKENHI (JP20K03828, JP 21H01036, JP 22H04461) and PRESTO (JPMJPR19L6).

*yoichi.nii.c1@tohoku.ac.jp

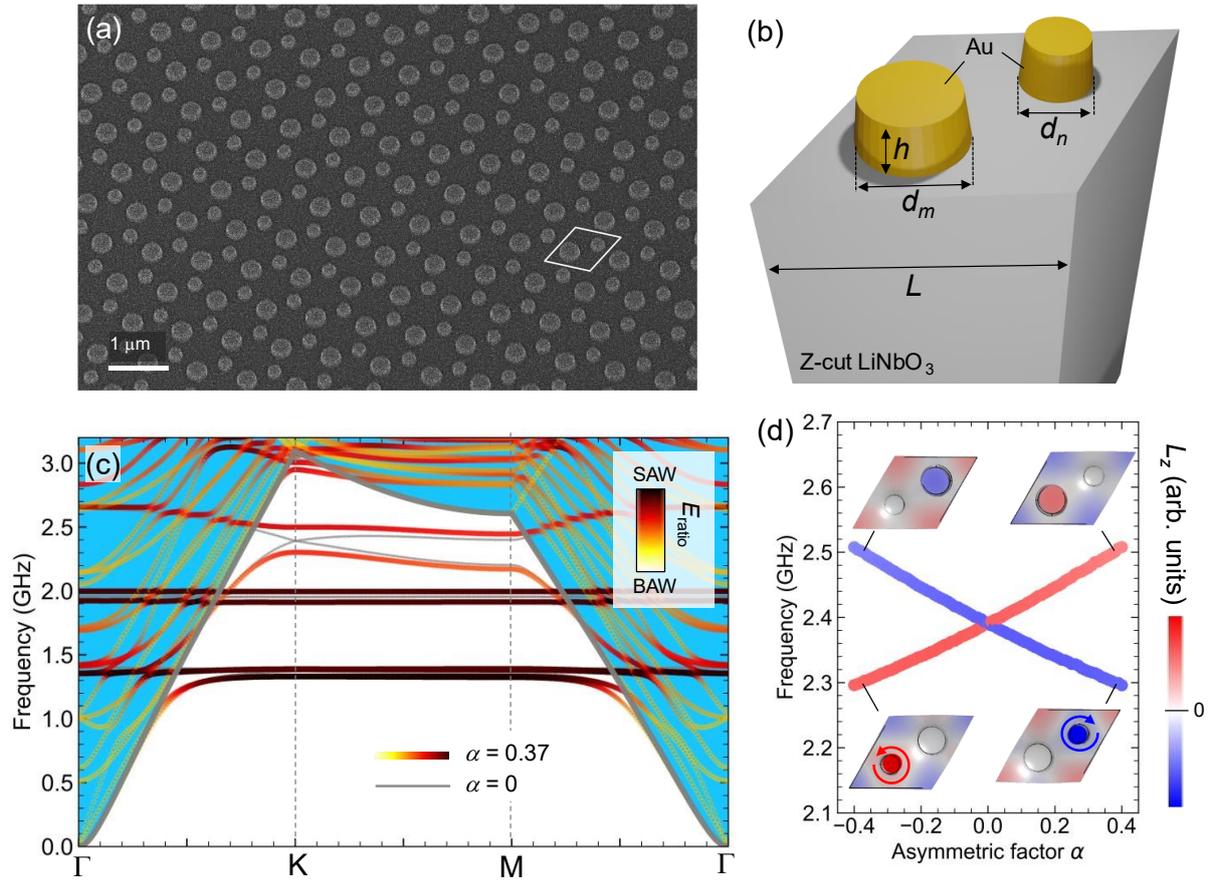

FIG. 1. (a) A scanning electron microscopy (SEM) image of a topological valley PnC with $\alpha = 0.37$. (b) A schematic diagram of the unit cell structure of the topological valley PnC in the band calculation. (c) Calculated band structure for $\alpha = 0.37$ and $\alpha = 0$. For the case of $\alpha = 0.37$, the color represents the ratio of elastic energy in the two pillars to that of the whole unit cell including the substrate, which represents the degree of surface (bulk) nature for the eigenmode. The blue shaded region represents the area above the sound line, where bulk acoustic modes inhabit. (d) Calculated frequencies of the upper and lower SAW modes at the K point as a function of asymmetric factor $\alpha$. Inset represents eigen modes of the lower and higher branches at $\alpha = \pm 0.37$. Red and blue colors correspond angular momentum along the Z-direction.

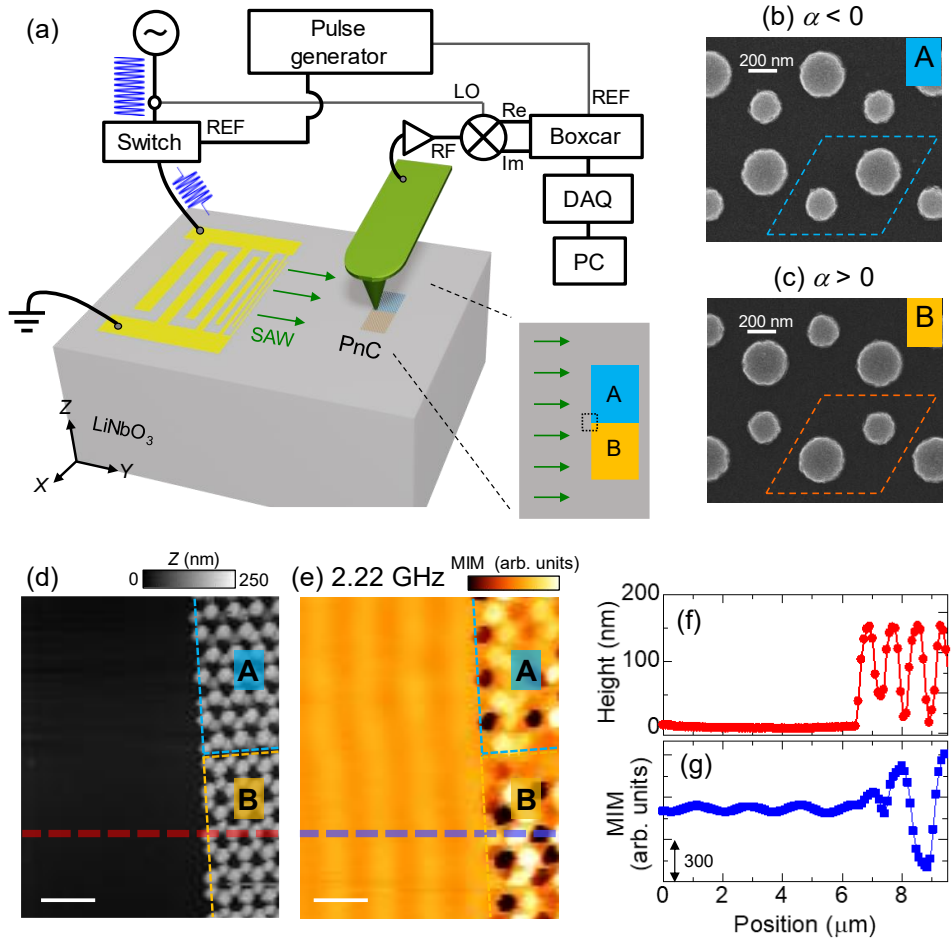

FIG. 2. (a) Setup of the MIM experiment. A chirped IDT and two mutually inverted honeycomb lattices (A, B lattices) made of metallic pillars were fabricated on a piezoelectric substrate. By using pulse-modulated microwaves and a time-domain gating technique, only the SAW signal was detected. X, Y, and Z axes correspond to the crystal orientation of $LiNbO_3$. A SAW propagates along Y-direction, which is parallel to the zigzag direction of the PnC. (b), (c) SEM images of two topological PnCs with (b) $\alpha = -0.37$ and (c) $\alpha = 0.37$. (d), (e) Examples of AFM and MIM images near the AB domain boundary, as shown in the inset of (a). During the scanning, a SAW with a frequency of 2.22 GHz was launched from the chirped IDT. Scale bars are 2 μm. (f), (g) Line profiles of (f) AFM and (g) MIM images shown in the red and the blue dot lines in (d) and (e), respectively.

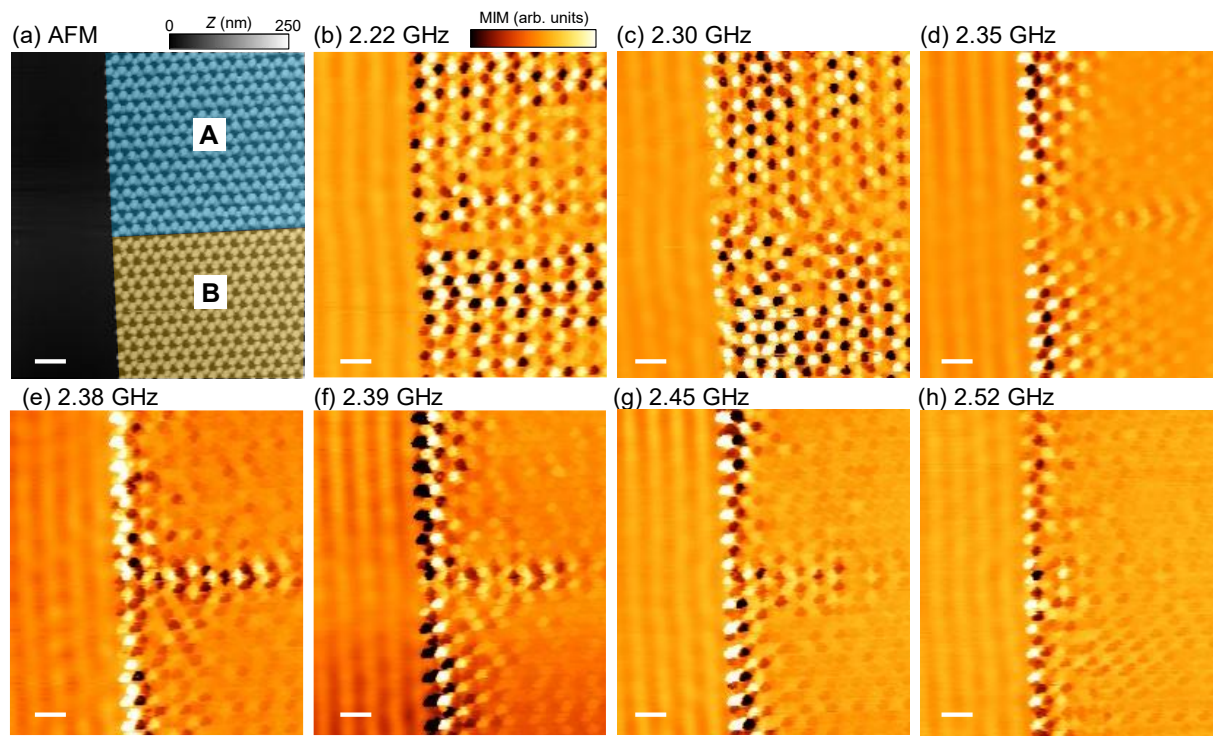

FIG. 3. (a) AFM image near the AB domain boundary. (b)-(h) MIM images at various SAW input frequencies. Scale bars are 2 μm.

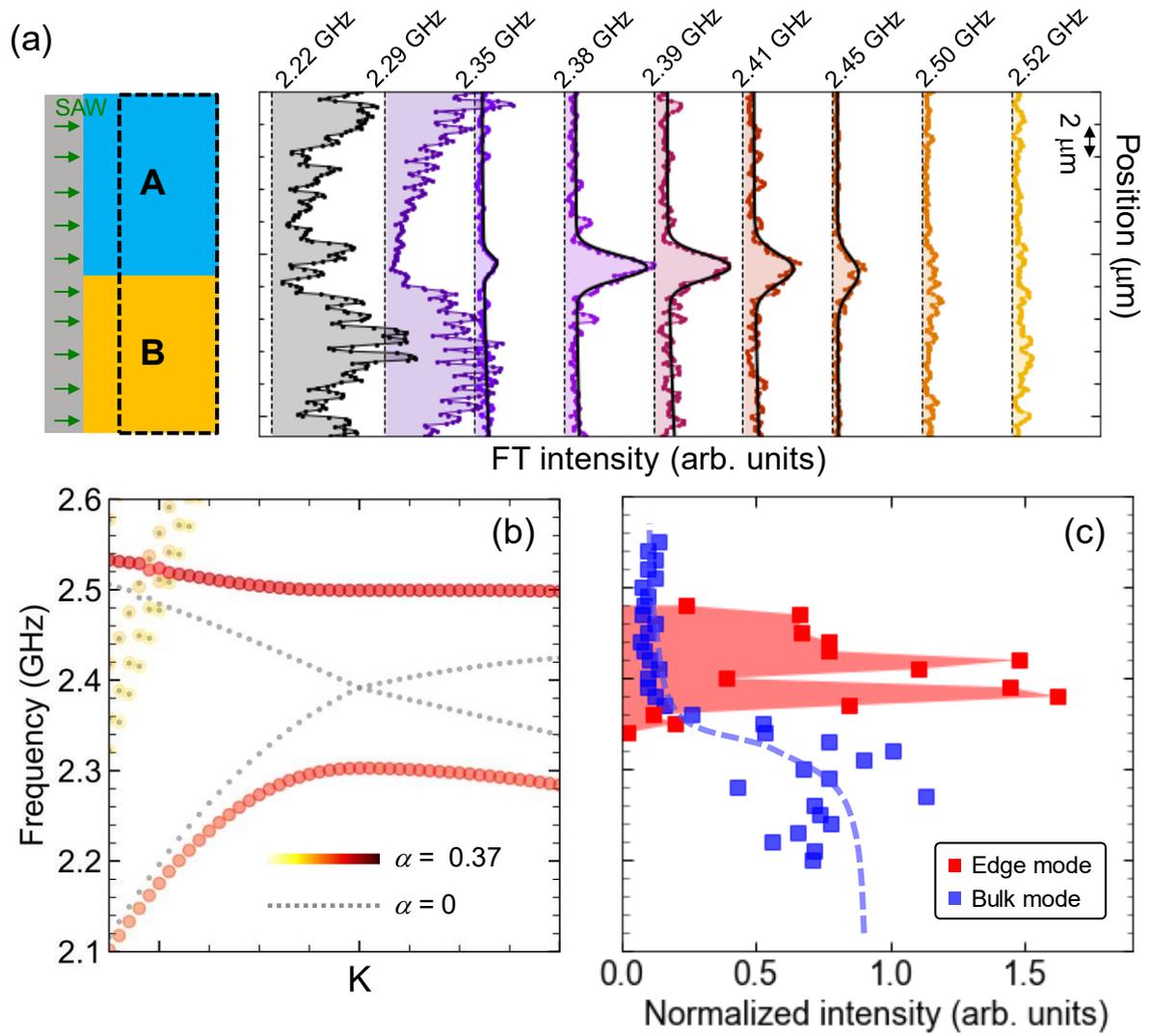

FIG. 4. (a) Vertical position dependence of SAW intensity around the A-B boundary at various frequencies. The black lines between 2.35 GHz and 2.45 GHz show the results of fitting to the Gaussian function for estimating the edge mode intensity. The dashed rectangle in the left inset represents the analyzed region for the estimation of intensity. (b) Acoustic band structure close to the SAW bandgap at the K point. (c) Edge and bulk mode intensities as a function of the excitation frequency. The blue dot line is a guide to the eye.

# Supplemental Material for

# Microwave impedance microscopy imaging of acoustic topological edge mode on a patterned substrate

Y. Nii,[1,2*] and Y. Onose[1]

*1. Institute for Materials Research, Tohoku University, Sendai 980-8577, Japan*

*2. PRESTO, Japan Science and Technology Agency (JST), Kawaguchi 332-0012, Japan*

### (A) Band calculation

The acoustic band calculation was performed using COMSOL Multiphysics 5.2. We deduced eigenfrequencies at each specific wavevector taking Floquet periodic boundary condition with the unit cell structure shown in Fig. 1(b). The physical parameters of LiNbO$_3$ substrate were density $\rho$ = 4.7 [g/cm$^3$], elastic stiffness tensors $C_{11}$ = 202.897 [GPa], $C_{12}$ = 52.9177 [GPa], $C_{13}$ = 74.9098 [GPa], $C_{14}$ = 8.99874 [GPa], $C_{33}$ = 243.075 [GPa], and $C_{44}$ = 59.9034 [GPa], relative permittivity coefficients $\varepsilon_{11}$ = 43.6, $\varepsilon_{33}$ = 29.16, piezoelectric constants represented by $e$-form $e_{15}$ = 3.69594 [C/m$^2$], $e_{22}$ = 2.53764 [C/m$^2$], $e_{31}$ = 0.193644 [C/m$^2$], and $e_{33}$ = 1.30863 [C/m$^2$]. The physical parameters of Au pillars were density $\rho$ = 19.3 [g/cm$^3$], Young's modulus = 79 [GPa], and Poisson's ratio = 0.4, respectively. For simplicity, we did not consider the Ti layer between Au pillars and LiNbO$_3$ substrate and instead assumed that pillars were made only of Au. We confirmed that the Ti adhesion layer hardly affected the acoustic band calculation. Since LiNbO$_3$ is anisotropic, the crystal orientation of LiNbO$_3$ was prescribed as in the experiment: Z is parallel to the surface normal and Y is parallel to Γ-K direction of PnC.

As discussed in the main text, some eigenmode is localized on the surface (SAW-like) and another mode is extended over the interior of substrate (BAW-like). In order to identify which eigenmode is SAW(BAW)-like, we have calculated a parameter $\eta$ of these eigenmodes given by

$$\eta = \iiint_{pillars} E(\boldsymbol{r})dV \;/\; \iiint_{unit\,cell} E(\boldsymbol{r})dV,$$

where $E(\boldsymbol{r})$ is the elastic energy density. $\eta$ takes the value from 0 to 1, the larger (smaller) value indicates more SAW-like (BAW-like). $\eta$ for each eigen mode of $\alpha$ = 0.37 was represented by color in Fig. 1(c).

In addition, we have evaluated $\alpha$-dependent angular momentum at the K point. As shown in the inset of Fig. 1(d), the blue and red colors represent the angular momentum along Z-direction, that was evaluated by $L_z(\boldsymbol{r}) = \mathrm{Re}[u(\boldsymbol{r})]\mathrm{Im}[v(\boldsymbol{r})] - \mathrm{Re}[v(\boldsymbol{r})]\mathrm{Im}[u(\boldsymbol{r})]$, in which the $u(\boldsymbol{r})$, $v(\boldsymbol{r})$, $w(\boldsymbol{r})$ are the X, Y, Z-components of the complex displacement vector $\boldsymbol{U}_k(\boldsymbol{r}, t) = (u(\boldsymbol{r}), v(\boldsymbol{r}), w(\boldsymbol{r})) \exp(i\omega t)$.

### (B) Device fabrication

To fabricate the elongated Au pillars in the PnC, trilayer resists based on MMA copolymer (EL6, Michrochem) and PMMA (495PMMA-A8, 950PMMA-A2 from Microchem) were spin-coated on LiNbO$_3$, and after the electron beam lithography and development, Ti (5 nm) and Au (153 nm) were evaporated by electron beams. Finally, PnC was obtained by the lift-off process. For the chirped IDT, we used a single layer resist of PMMA (950PMMA-A2), and Ti (5 nm)/ Au (60 nm) were made by electron beam evaporation. The length and the total number of IDT fingers are 126 μm and 200, respectively. The designed width of the fingers $d_0$ varies from 371 nm to 404 nm. The space between the fingers is also $d_0$.

### (C) MIM imaging

For the MIM measurement, we used a commercial conductive cantilever (12Pt-400B, Rocky Mountain Nanotechnology), that has a low Young's modulus of 0.3 N/m. It prevents damage to the PnC during scanning. Microwave coaxial cable was directly connected to the cantilever by wire-bonding without inserting an impedance matching unit typically involved in the MIM system [26,29-31]. Although there is an impedance mismatch between the coaxial cable and the cantilever tip, we could cover wide frequency spectrum.

Similar to the previous MIM systems [26,29-31], we could obtain two orthogonal MIM images, that are referred to as MIM-Re and MIM-Im, respectively. Figure S1 presents typical MIM-Re and MIM-Im images. These have a π/2 phase difference, but other features are quite similar. Therefore, we show only MIM-Re images.

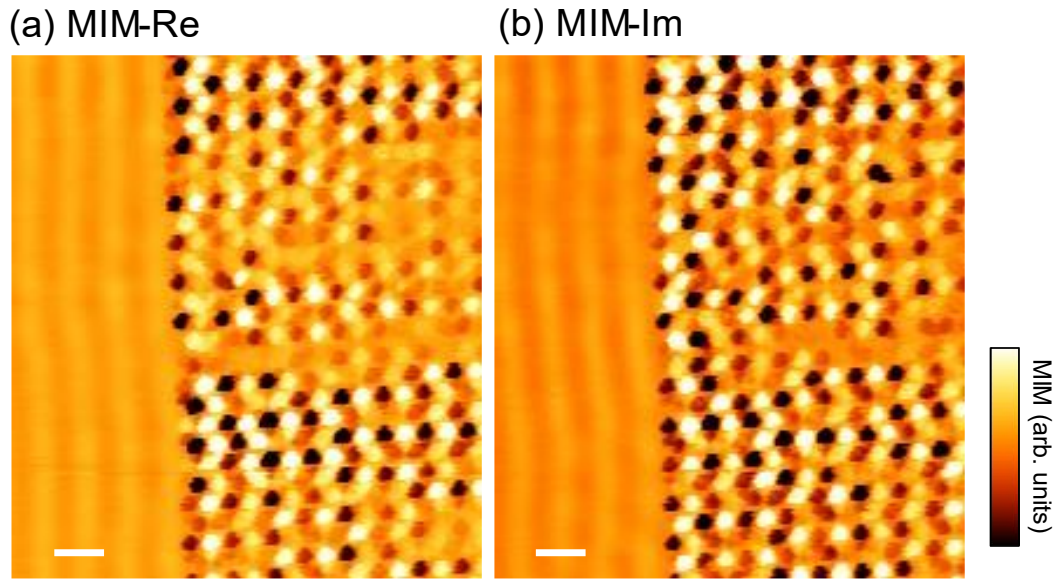

FIG. S1. (a) MIM-Re and (b) MIM-Im images at a SAW frequency of 2.22 GHz. The observed region is identical to that shown in Fig. 3. Scale bars are 2 μm.

### (D) Fourier analysis

Here we explain the Fourier transformation (FT) analysis of MIM images used for obtaining the quantities plotted in Fig. 4(a). First, we focused on the right half of the PnC as shown in the left schematic of Fig. 4(a). Then, one-dimensional FT was carried out for each MIM image along the horizontal direction parallel to the SAW propagation direction (*i.e.*, Γ-K direction). While the uniform SAW gives rise to a peak around the wave vector ranging from 0.47 μm$^{-1}$ to 0.74 μm$^{-1}$ in the FT profile, extrinsic noise may contribute to the amplitude over the much wider momentum region. To precisely estimate the intrinsic SAW intensity, we integrated FT profiles in the wavevector region from 0.38 μm$^{-1}$ to 0.95 μm$^{-1}$. Note that K point corresponds to $2/(3L) = 0.79$ μm$^{-1}$.

To evaluate the frequency dependence of bulk SAW mode transmission into the PnC, we have performed a similar analysis. To avoid the effect of A-B boundary, we focused on the lower half of the B lattice as shown in Fig. S2(a). We did the MIM measurement at various frequencies as exemplified in Figs. S2(b)-(d), and performed FT procedure. Figure S2(e) shows the vertical position dependence of the FT intensity at various frequencies. In Fig. 4(c), we plot the averaged FT intensity inside the PnC normalized by that outside the PnC.

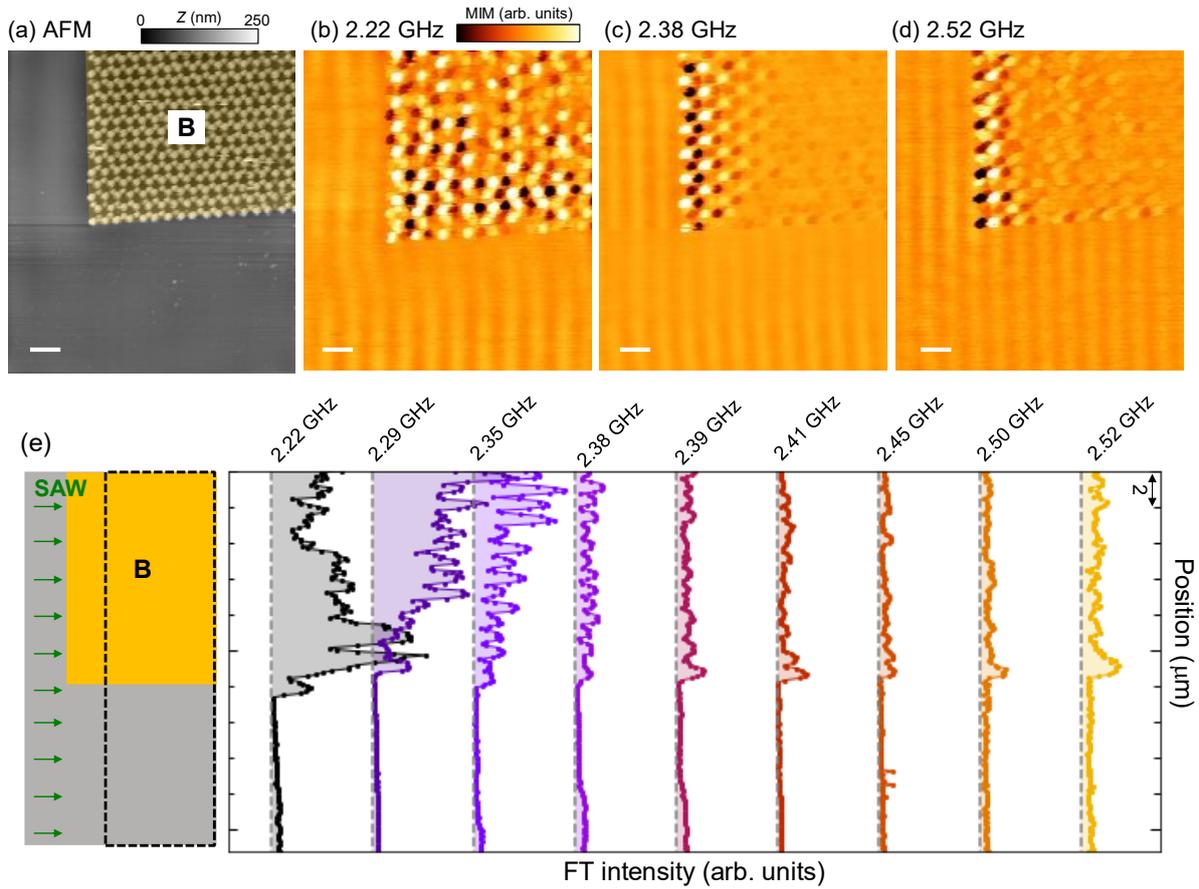

FIG. S2. (a) AFM image near the corner of the PnC. (b)-(d) MIM images at various SAW input frequencies. Scale bars are 2 μm. (e) Vertical position dependence of FT intensities at various frequencies. The left panel schematically represents the analyzed region, in which the dot rectangular represents the area used for the FT analysis.